# Enhancing visibility of graphene on arbitrary substrates by microdroplet condensation

Hugo Gonçalves[1], Michael Belsley[1], Cacilda Moura[1], Tobias Stauber[2] and Peter Schellenberg[1]*

[1]Centro de Física, University of Minho, Campus de Gualtar, Pt - 4710-057 Braga, Portugal

[2]Department of Condensed Matter Physics, University Autónoma of Madrid Campus of Cantoblanco, E-28049 Madrid, Spain

## Abstract

In order to take advantage of the enormous potential of graphene for future electronic micro-circuits and other applications it is necessary to develop reliable, rapid and widely applicable methods to visualize graphene based structures. We report here on a micro-droplet condensation technique, which allows for quick visual identification of graphene on a variety of substrates, including some which were previously considered unsuitable for the visualization of carbon layers. The technique should also be applicable to visualize artificially patterned graphene structures which are expected to be key technologically enabling components in electronic micro-circuits and other applications.

* to whom correspondence should be addressed, peter.schellenberg@fisica.uminho.pt

Monolayer [1-3], bilayer [4] and few layer graphene [5] have attracted intense research interest in recent years [6,7] due to their unique electronic transport [8] and elastic [9] properties which allow one to probe a variety of fundamental questions in physics [10] as well as to enable the potential development of distinctive nanoelectronic devices [11]. While a multitude of mostly optics based methods have been proposed to identify single and few atomic layer samples[12], the process is a time consuming and tedious task. For example, interference enhanced reflection contrast due to refractive index differences between a dielectric layer on a silicon wafer and the graphene flakes is often used [13-16]. Other imaging methods are based on Raman and Rayleigh scattering [17,18], on fluorescence quenching of dyes by graphene layers [19], or on ellipsometry [20]. All of these methods require complex equipment or an optimized substrate coating. Here, we report on a generally applicable and easy to use technique for the identification of potential graphene flakes by microdroplet condensation. The method is already well-known in micropatterning for the visualization of surface features of different hydrophobicities. In particular, this technique is suitable for making graphene visible in a standard reflection microscope on virtually any surface including plastics or uncoated metals, for which no method was previously available [12]. Graphene can also be identified on transparent surfaces in transmission, which is difficult to achieve otherwise, as the absorption of a graphene monolayer is just 2.3 % of white light [21].

The technique may also be utilized to analyze artificially produced graphene based micropatterns, such as microelectronic or microoptical devices as envisioned in recent developments of graphene applications. Assessing the quality and uniformity of large-area graphene films [22,23] for use in transparent conductor applications such as flat screens or solar panels may also become possible.

Upon exhaling onto a simple glass surface, the resulting breath condensation shows the presence of fatty deposits, due to the difference in hydrophobicity. This observation was already described by Rayleigh [24], and was later exploited to investigate microscopic heterogeneities on surfaces, which resulted in optical condensation images [25-28]. The method is capable of visualizing heterogeneities in self-assembled monolayers [27,29] or natural [30] or artificial [25,31,32] micropatterning of a surface. Regions of different hydrophobicity result in the appearance of condensation features and the spatial resolution of this patterning depends on the size of the condensed microdroplets. In this letter, we show that microdroplet condensation can be easily applied to readily show single and few-layer graphene flakes on almost any kind of surface, which would otherwise be difficult to spot.

Apparently, graphene has an especially high hydrophobicity, making it easy to identify graphene flakes, even when the optical contrast in relation to the substrate is low or vanishing. The method is particularly suited for identifying flakes with spatial dimensions of 1 µm or larger, the resolution depending on the substrate and the vapor used. We have tested a variety of substrates, including microscope glass slides, adhesion slides (Menzel – 'Polysine' and -'Plus') as well as silicon and glass wafers spin-coated with a layer of polymethylmethacrylate (PMMA) [14] or polystyrene (PS). Metal surfaces such as polished stainless steel and uncoated silicon were also used. These materials cover a wide range of hydrophobicities and other surface properties, demonstrating the generality of the technique.

In the experiments, the flakes were deposited by mechanical cleavage of 2 - 5 mm large natural graphite pieces (NGS Naturgraphit 'graphenium '). The resulting monolayers were typically 25 – 500 µm$^2$ in size. Subsequently, the polymer coated substrates were cleaned with ethanol and the uncoated surfaces were cleaned with chloroform to remove glue stains and other contaminations. The droplet condensation figures are produced by blowing water [28], Ethyleneglycol (EG) or Diethyleneglycol (DEG) [32] vapor over the substrate with the graphene flakes immobilized on its surface.

In principle, water droplet condensation figures could be produced by simply exhaling over the slide surface. To allow for longer viewing times it is helpful to cool the surface using a peltier element attached below the sample [31]. In order to better control the droplet size and also to permit the use of different vapors, an apparatus adapted from Schäfle et al [32] was employed to generate the vapor. In short, nitrogen gas was bubbled through a gas wash bottle filled with the respective liquid. The gas flowed through the outlet of a glass tube topped with a fritted glass piece for bubble dispersion, and the vapor exited via a nozzle. Water vapor was generated at room temperature and blown directly over the slide placed on the microscope stage. EG and DEG were heated in the same setting to around 70° and 120° C respectively. The slides were fixed at a distance of 40 mm in front of a 5 mm wide nozzle covered with a sheet of lens tissue to prevent large droplets from exiting and exposed to the vapor for 10 s, after which the slides were transferred to the microscope where the droplets remained stable for several minutes. The observation time could be extended by using a peltier element to induce moderate cooling. While DEG was used in the original publication [32], the usage of EG turned out to be the superior choice in our experiments, as the droplet sizes were more homogeneous and the high DEG vapor temperature of ~120° C occasionally damaged the graphene flakes.

The droplets typically evaporated, leaving no residue, but the surface could also be cleaned with an appropriate solvent. The investigations were performed using a Nikon Optiphot metallurgical reflection microscope.

For an illustration of the technique, Fig. 1a shows a graphene flake, which consists of a few layers of carbon to make it at least faintly visible in the photograph prior to droplet condensation, and the contrast enhancement due to water (Fig. 1b) or EG (Fig. 1c) droplet condensation. Note that the presence of the droplets not only delimits the area of the flake, but also helps to increase the visual contrast by reducing the amount of light collected from the areas covered by the droplets. Fig. 2 and 3 illustrate the identification of graphene flakes on glass and metal substrates respectively, that were virtually invisible prior to droplet condensation and which were subsequently confirmed by Raman microscopy to be monolayers.

Generally, the droplet size is the limiting factor for the spatial resolution. This in turn depends on the chemical used for vapor generation, on the details of the vapor preparation and on the substrate surface. Resolutions of roughly 2.5 µm can be obtained for water on glass and PMMA and 1 µm for water on polysine surfaces. The best resolution (0.7 µm) was achieved by depositing EG on Menzel slide polysine and 'plus' surfaces. Although it is not possible to distinguish graphene monolayers from few-layer sheets by this method, it does allow a quick and easy identification of candidate flakes on the slide, which are imperceptible or barely visible by eye. For an unambiguous determination of the number of layers, we subsequently employed Raman microscopy to investigate each candidate flake separately.

To conclude, we were able to show that micro-droplet condensation can be employed to quickly and easily visualize graphene and few-layer graphite on a wide variety of substrates without the need for any specific surface modification or preparation. There is no other currently available method that is capable of readily visualizing graphene flakes on plastic surfaces [12] and metals. The technique will not only facilitate the identification of graphene flakes but may also have a great potential for the quick and non-invasive visualization of artificially patterned structures in graphene-based devices such as electronic micro-circuits, micro-optical settings and others.

We thank J. A. Santos for instrumental support and W. Fritzsche, K. S. Novoselov and N. M. R. Peres for useful discussions. This work has been supported by FCT under grant PTDC/FIS/101434/2008 and by MIC under grant FIS2010-21883-C02-02.  H.G. received support from FCT, grant no. BII/UNI/0607/FIS/2009.

# Figures

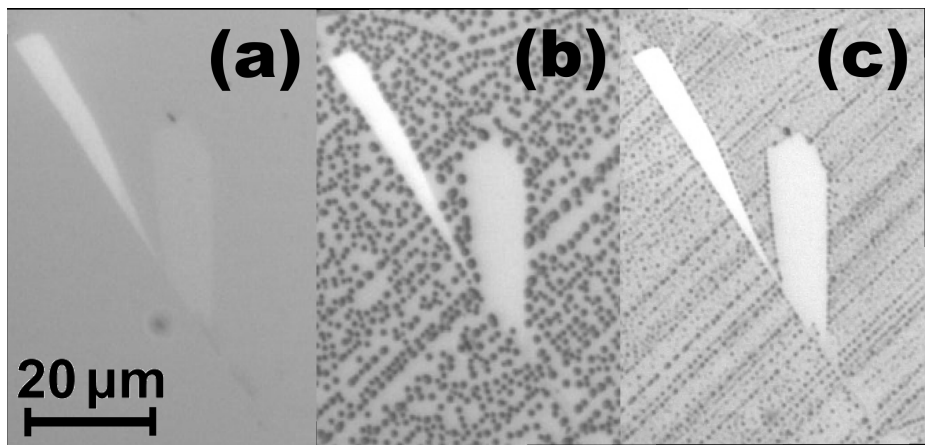

**Figure 1. Few-layer flake tested with different droplet condensates**

(a): A several layer graphene flake on a Menzel 'plus' adhesion microscopy slide. The flake is barely visible by eye or by camera. (b): The same flake observed using water droplet condensation. Note that the intensity increase of the flake is due to the darkening of the environment around the flake. (c): The same flake observed using Ethylene Glycol (EG) droplet condensation. The ruler scale is 20 µm.

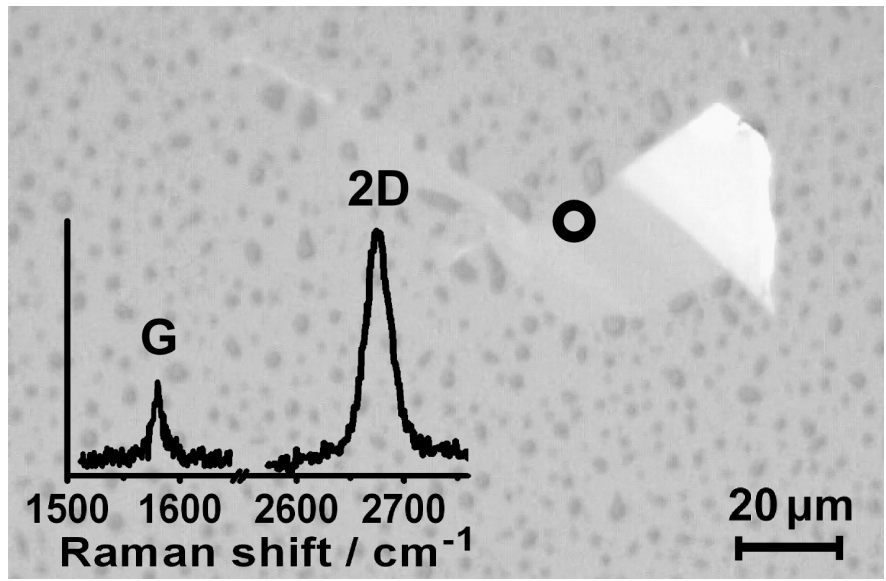

**Figure 2. Monolayer flake with droplet condensation**

A monolayer graphene flake on a BK7 glass microscopy slide with Diethylene Glycol (DEG) droplet condensation. Parts of the flake were confirmed by Raman spectroscopy to be a monolayer as displayed in the inset for the indicated region. The intensity ratio between 2D and G line and the shape of the 2D line in the Raman spectrum are indicative of a graphene monolayer.

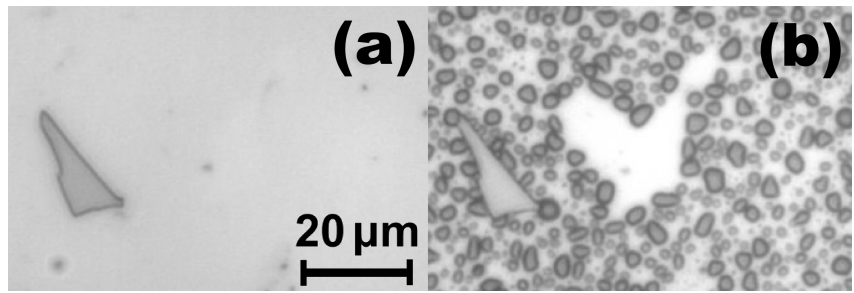

**Figure 3.** **Monolayer flake on silicon metal**

(a): A flake is just faintly visible on a silicon metal surface (without any dielectric coating apart from the natural oxidation layer). (b): The image after water droplet condensation. A part of the flake becomes visible which can not be observed optically without droplet condensation even by close inspection of the area in question. It has been shown to be a monolayer by Raman microscopy.